\documentclass[
twocolumn,
showpacs,preprintnumbers,
bibnotes,
 amsmath,amssymb,
 aps,
 pra,
superscriptaddress,
longbibliography,
]{revtex4-1}

\usepackage[version=3]{mhchem} 
\usepackage[dvipsnames]{xcolor}
\usepackage{textcomp}
\usepackage{graphicx}
\usepackage{amsmath}
\usepackage{fixmath}
\usepackage{amsfonts}
\usepackage{amssymb}

\newcommand{\micron}{\hbox{\textmu}\text{m}}
\newcommand{\figurewidth}{\columnwidth}

\begin{document}

\title{
Valley polarization of trions in monolayer MoSe$_2$ interfaced with bismuth iron garnet 
}

\author{Vasily Kravtsov}
\email{Corresponding author: vasily.kravtsov@metalab.ifmo.ru}
\affiliation{Department of Physics and Engineering, ITMO University, Saint Petersburg 197101, Russia}
\author{Tatiana Ivanova}
\affiliation{Department of Physics and Engineering, ITMO University, Saint Petersburg 197101, Russia}
\author{Artem N. Abramov}
\affiliation{Department of Physics and Engineering, ITMO University, Saint Petersburg 197101, Russia}
\author{Polina V. Shilina}
\affiliation{Center for Photonics and 2D Materials, Moscow Institute of Physics and Technology, National Research University, Dolgoprudny 141700, Russia}
\author{Pavel O. Kapralov}
\affiliation{Russian Quantum Center, Skolkovo Innovation City, Moscow 121353, Russia}
\author{Dmitry N. Krizhanovskii}
\affiliation{Department of Physics and Engineering, ITMO University, Saint Petersburg 197101, Russia}
\affiliation{Department of Physics and Astronomy, University of Sheffield, Sheffield S3 7RH, UK}
\author{Vladimir N. Berzhansky}
\affiliation{Vernadsky Crimean Federal University, Simferopol 295007, Russia}
\author{Vladimir I. Belotelov}
\affiliation{Russian Quantum Center, Skolkovo Innovation City, Moscow 121353, Russia}
\affiliation{Vernadsky Crimean Federal University, Simferopol 295007, Russia}
\affiliation{Faculty of Physics, Lomonosov Moscow State University, Moscow 119991, Russia}
\author{Ivan A. Shelykh}
\affiliation{Science Institute, University of Iceland, Dunhagi-3, IS-107 Reykjavik, Iceland}
\affiliation{Department of Physics and Engineering, ITMO University, Saint Petersburg 197101, Russia}
\author{Alexander I. Chernov}
\affiliation{Center for Photonics and 2D Materials, Moscow Institute of Physics and Technology, National Research University, Dolgoprudny 141700, Russia}
\affiliation{Russian Quantum Center, Skolkovo Innovation City, Moscow 121353, Russia}
\affiliation{NTI Center for Quantum Communications, National University of Science and Technology MISiS, Moscow 125009, Russia}
\author{Ivan V. Iorsh}
\email{Corresponding author: i.iorsh@metalab.ifmo.ru}
\affiliation{Department of Physics and Engineering, ITMO University, Saint Petersburg 197101, Russia}

\date{\today}



\begin{abstract}
\noindent 
\textbf{
Interfacing atomically thin van der Waals semiconductors with magnetic substrates enables additional control on their intrinsic valley degree of freedom and provides a promising platform for the development of novel valleytronic devices for information processing and storage.
Here we study circularly polarized photoluminescence in heterostructures of monolayer MoSe$_2$ and thin films of ferrimagnetic bismuth iron garnet.
We observe strong emission from charged excitons with negative valley polarization, which switches sign with increasing temperature, and demonstrate contrasting response to left and right circularly polarized excitation, associated with finite out-of-plane magnetization in the substrate.
We propose a theoretical model accounting for magnetization-induced imbalance of charge carriers in the two valleys of MoSe$_2$, as well as for valley-switching scattering from B to A excitons and fast formation of trions with extended valley relaxation times, which shows excellent agreement with the experimental data.
Our results provide new insights into valley physics in 2D semiconductors interfaced with magnetic substrates.
}
\end{abstract}

\maketitle

\noindent
Transition metal dichalcogenides (TMDs) in their monolayer form are attractive candidates for integration into future on-chip opto-electronic devices as atomically thin and air-stable materials that support strongly bound excitons~\cite{Mak2016, Wang2018, Akinwande2019}.
Broken inversion symmetry and strong spin-orbit coupling in these systems result in the emergence of two energy-degenerate but nonequivalent valleys K/K' coupled to photons of opposite helicity~\cite{Yao2008, Mak2012}, which gives access to a new ``valley" degree of freedom and provides opportunities for the development of TMD-based valleytronic devices~\cite{Vitale2018, Schaibley2016}.


A promising general approach to further control of the TMD excitonic valley properties is via interfacing with magnetic materials~\cite{Zhang2016}.
It allows, for example, lifting the energy degeneracy at the K and K' points through proximity effects and eliminating the need for large external magnetic fields~\cite{Zhao2017} as required for potential information storage applications.
This has generated much interest in experimental studies of TMD systems put in proximity with various bulk, thin films, or layered van der Waals ferromagnetic and antiferromagnetic  crystals~\cite{Norden2019, Huo2020, Zhang2019, Zhong2017, Lyons2020, Ciorciaro2020}.

Iron garnets, in particular yttrium iron garnet (YIG) ~\cite{Ng2020, Tsai2021}, have recently attracted attention as promising magnetic substrates for TMD-based heterostructures.
Besides the unique spin properties of YIG enabling, for example, efficient magnon-exciton coupling~\cite{Gloppe2020}, this ferrimagnetic insulator with low surface roughness can significantly dope adjacent TMD monolayers, thus boosting the oscillator strength of trions without significantly changing the quality of their optical response~\cite{Peng2017}.
This provides an opportunity to study valley physics of charged excitons in TMDs, as it was discussed in recent reports of trion valley polarization in MoS$_2$/YIG~\cite{Peng2017} and WS$_2$/YIG samples~\cite{Carmiggelt2020}.

\begin{figure*}[t]
	\includegraphics[width=0.8\textwidth]{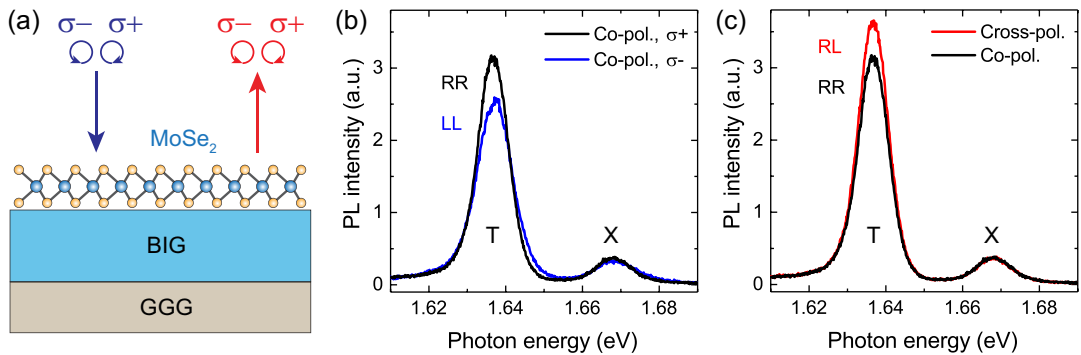}
	\caption{\textbf{Measurement of circularly polarized photoluminescence in MoSe$_2$/BIG structures.} (a) Schematic of the sample, consisting of monolayer MoSe$_2$ transferred onto thin film of bismuth iron garnet (BIG) on gadolinium gallium garnet (GGG) substrate, and different excitation/detection circular polarization channels. (b) PL spectra measured in a co-polarized detection channel with right (RR) and left (LL) circularly polarized laser excitation. (c) PL spectra measured in co- (RR) and cross- (RL) polarized detection channels with right circularly polarized laser excitation.}
	\label{fig:Setup}
\end{figure*}

Here we study circularly polarized photoluminescence of trions in MoSe$_2$ monolayers interfaced with thin films of a similar magnetic compound from the iron garnet family, bismuth iron garnet (BIG)~\cite{Ignatyeva2020} deposited on gadolinium gallium garnet (GGG) substrates.
We observe contrast in co-polarized PL for left- and right- circularly polarized excitation, associated with a finite energy splitting between K and K' points in MoSe$_2$ induced by the BIG film proximity, and measure its temperature dependence.
In addition, we report enhanced circular polarization of trions opposite to that of the excitation (negative polarization), which exhibits a crossover to a finite positive polarization with increasing temperature.
To explain these results, we develop a theoretical model based on intervalley scattering between B and A excitons.
The validity of our model is supported by experimental data of the dependence of valley polarization on the excitation laser wavelength. Our findings contribute to the understanding of the valley physics in TMD-based systems and highlight TMD/BIG structures as a promising platform for studying charged excitons and their valley dynamics.




In the experiment, we measure circular-polarization-resolved photoluminescence spectra of monolayer MoSe$_2$ interfaced with thin (3 \textmu m) bismuth iron garnet films on 500 \textmu m thick GGG (Gd$_3$Ga$_5$O$_{12}$) substrates as illustrated in Fig.~\ref{fig:Setup}a.
Magnetic films were grown by liquid phase epitaxy.
Specifically, two types of the garnet composition were used: (BiY)$_3$(FeScAl)$_5$O$_{12}$ with in-plane magnetic anisotropy and (BiLuEu)$_3$(FeGaAl)$_5$O$_{12}$ with an easy axis along the out-of-plane direction.
High-quality flakes of MoSe$_2$ monolayers were mechanically exfoliated from a commercial bulk crystal (HQ Graphene) with Nitto tape.
TMD/BIG structures were assembled via dry transfer of MoSe$_2$ onto BIG films with viscoelastic polydimethylsiloxane stamps.
We performed photoluminescence (PL) hyperspectral mapping on the prepared samples to determine areas with uniform contact between TMD and substrate and carried out further measurements in the selected locations.

Spectroscopic measurements were performed with samples mounted in a closed-cycle ultralow-vibration cryostat (Advanced Research Systems) and maintained at a controlled temperature in 6-50~K range.
Samples were excited either with a 632~nm (1.96~eV) cw HeNe laser or with light from  a supercontinuum source (Fianium WhiteLase) spectrally filtered down to 4~nm bandwidth with center wavelength continuously tuned from 550~nm to 750~nm.
The excitation laser beam was focused onto the sample with a long-working distance microscope objective (Mitutoyo 50X/0.42) yielding a spot size of $\sim 1$~\micron, the incident power was kept at $\sim 1$~\textmu W.
PL spectra were collected via the same microscope objective and detected with a spectrometer and liquid nitrogen cooled CCD camera (Princeton Instruments SP2500 + PyLoN).
Right and left circular polarization states were selected in the excitation and detection channels with a combination of broadband polarizers and superachromatic $\lambda/4$ and $\lambda/2$ plates.

Typical PL spectra measured on MoSe$_2$/BIG samples with an easy axis (out-of-plane magnetization) in different polarization channels at 6~K excited by 633~nm cw laser are shown in Fig.~\ref{fig:Setup}b,c (see Supplementary Information for the results on the samples with in-plane anisotropy).
The two peaks correspond to emission from the neutral (X) and charged (T, trion) exciton complexes.
In comparison to the case of monolayer MoSe$_2$ on Ta$_2$O$_5$ substrates (see Supplementary Information), the trion PL is enhanced while the neutral exciton PL is significantly suppressed, which is due to electron doping from BIG films.

Spectra with right ($\sigma_{+}$, R) and left ($\sigma_{-}$, L) circularly polarized excitation detected in the co-polarized channel (RR, LL) are shown in Fig.~\ref{fig:Setup}(b).
To quantify the observed contrast, we introduce a parameter $\rho = (I_\mathrm{RR} - I_\mathrm{LL})/(I_\mathrm{RR} + I_\mathrm{LL})$, where $I_\mathrm{RR}$, $I_\mathrm{LL}$ are PL intensities in the co-polarized channel when excited with right and left circularly polarized light, respectively.
The parameter $\rho$ is related to the imbalance of carrier population in the monolayer MoSe$_2$ K/K' valleys due to Zeeman splitting in the proximity of the BIG film, leading to the corresponding but opposite imbalance of trion population, and is estimated as $\sim 10\%$ for the trion PL peak.

Spectra with right circularly polarized (R) excitation and different detection polarizations are shown in Fig.~\ref{fig:Setup}(c), where RR corresponds to co-polarized PL, and RL to cross-polarized.
We quantify the corresponding contrast with the degree of circular polarization (DOCP) defined as $\mathrm{DOCP} = (I_\mathrm{co} - I_\mathrm{cross})/(I_\mathrm{co} + I_\mathrm{cross})$, where $I_\mathrm{co}$, $I_\mathrm{cross}$ are PL intensities in the co- and cross-polarized channels, respectively.
The DOCP value characterizes the degree of valley polarization in the two valleys K/K' of the MoSe$_2$ monolayer and is estimated for the trion PL peak as $\sim -7\%$.
The negative DOCP value implies that, following excitation in a certain valley, the resulting population of trions becomes higher in the opposite valley.

\begin{figure}[t]
	\includegraphics[width=\figurewidth]{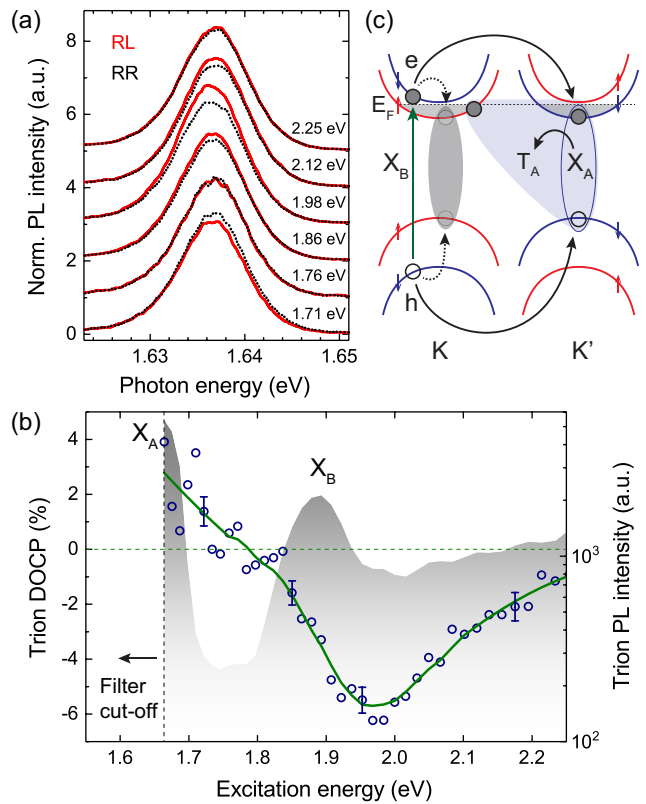}
	\caption{\textbf{Excitation frequency dependence of the trion PL response in MoSe$_2$/BIG structures.} (a) Normalized PL spectra taken in co- (RR, black curves) and cross- (RL, red curves) polarized detection channels for selected laser excitation frequencies indicated on the right. (b) Measured degree of circular polarization (DOCP) of the trion PL peak (blue circles, the green line is a guide to the eye, left vertical axis) and corresponding PL intensity (shaded gray curve, right vertical axis) as functions of the excitation photon energy. X$_\mathrm{A}$ and X$_\mathrm{B}$ indicate peaks corresponding to excitation of A and B excitons. (c)	Schematic of the process resulting in the negative valley polarization of trions.}
	\label{fig:Wavelength}
\end{figure}

To clarify the mechanism of the negative trion DOCP, we measured PL spectra in the co- and cross-polarized channels for different excitation wavelengths.
The obtained spectra for selected excitation wavelengths are shown in Fig.~\ref{fig:Wavelength}a, where black and red curves correspond to co- and cross-polarized PL, respectively.
For low excitation photon energies $\hbar\omega_\mathrm{exc} \sim 1.7$~eV close to the trion emission energy, we observe a small positive DOCP ($3-4 \%$) of the trion PL peak, which switches sign and grows negative for higher photon energies $\hbar\omega_\mathrm{exc} \sim 2.0$~eV and ultimately approaches zero for $\hbar\omega_\mathrm{exc} > 2.2$~eV.

Fig.~\ref{fig:Wavelength}b shows the extracted DOCP for the trion PL peak as a function of the excitation photon energy (circles and green curve as a guide to the eye, left axis), together with the corresponding trion PL excitation spectrum (gray, right axis).
The PL excitation spectrum exhibits 2 peaks at $\sim 1.9$~eV and $< 1.70$~eV, corresponding to the B and A excitons, respectively, in monolayer MoSe$_2$.
As seen in the figure, trion DOCP values are negative when exciting at or slightly higher than the B exciton frequency and become positive when exciting close to the A exciton frequency, which we will address in the discussion below.

\begin{figure}[t]
	\includegraphics[width=\figurewidth]{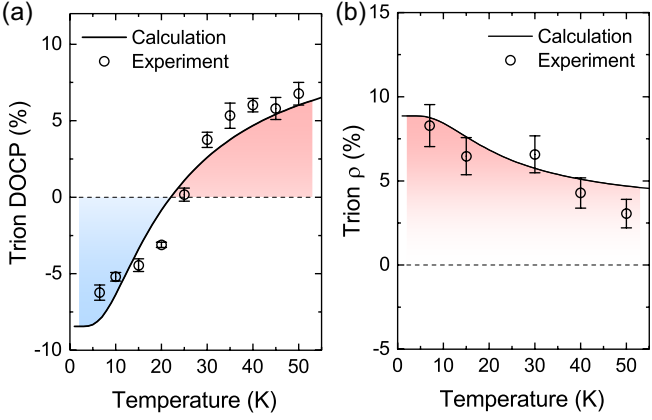}
	\caption{\textbf{Temperature dependence of the trion PL circular polarization in MoSe$_2$/BIG structures.} (a) Temperature dependence of the degree of circular polarization. (b) Temperature dependence of the parameter $\rho$ defined as contrast between co-polarized PL response for right and left circularly polarized excitation (see text). Experimental data points are shown with circles, and results of calculations according to the model are shown with shaded curves.}
	\label{fig:Temperature}
\end{figure}

We further investigate the mechanisms behind the observed parameters DOCP and $\rho$ by measuring their dependencies on temperature.
In Fig.~\ref{fig:Temperature} we show the values of DOCP (a, circles) and $\rho$ (b, circles) obtained experimentally for the trion PL peak at varying sample temperatures.
As seen in the plots, the trion DOCP changes sign at $T \sim 25$~K and becomes positive for higher temperatures, while the parameter $\rho$ exhibits only a slight decrease with increasing temperature.


In the following discussion, we address the observed trends and develop a theoretical model that explains the experimental findings.
The two distinct features observed in the experiment are (i) the negative degree of trion circular polarization at low temperatures, which changes sign as temperature is elevated, and (ii) the contrast between co-polarized trion PL response for right and left circularly polarized pump.

The former effect does not require time reversal symmetry breaking in the sample and thus can also be observed, although less pronounced, in monolayer MoSe$_2$ on substrates without out-of-plane magnetization when excited close to resonance with B exciton (see Supplementary Information and Ref.~\cite{Zhang2015}).
We explain this effect by the presence of two competing energy relaxation pathways for the B exciton in MoSe$_2$ as illustrated in Fig.~\ref{fig:Wavelength}c: the first one involves spin-preserving phonon assisted scattering of the B exciton in the $K(K')$ valley to the opposite $K'(K)$ valley (shown with solid arrows); the second one corresponds to the exchange driven intravalley spin-flip process~~\cite{guo2019exchange} (shown with dashed arrows).
Both of these processes are irreversible due to the large energy difference between A and B excitons.
At low temperatures, the intravalley process is suppressed since excitons mostly occupy the band minima, and the intravalley exchange coupling is proportional to the mean thermal momentum of the excitons~\cite{yu2014valley}.
Thus the intervalley relaxation with the emission of a phonon dominates, which results in an excess population of A excitons in the opposite valley.
While neutral excitons in monolayer MoSe$_2$ rapidly depolarize~\cite{Wang2015}, trions exhibit extended valley relaxation times and thus partially retain the negative valley polarization.
As temperature is increased, the intravalley process becomes dominant, thus switching the sign of the trion valley polarization.
We note that valley-switching processes involving scattering between A and B excitons in other TMD compounds have been discussed in literature previously~\cite{Manca2017,Berghauser2018}; however, negative DOCP was reported only for the case of upconverted emission.
The dynamical processes underlying the formation of trions with negative valley polarization in TMD-based heterostructures can be in future studied with time-resolved spectroscopy using, for example, time-resolved Kerr rotation~\cite{Kravtsov2021}.

The second observed effect, namely the difference between RR and LL signals, requires the breaking of time reversal symmetry in the sample and can be observed only in the MoSe$_2$/BIG sample with out-of-plane magnetization. We associate the observed contrast with the difference in the residual electron concentrations induced by the electron Zeeman effect. We note however that Zeeman splitting is not observed directly as its value is below our spectral resolution.
The trion oscillator strength $f_\mathrm{T} \propto n_e f_\mathrm{X}$ is proportional to the residual electron concentration and the neutral exciton oscillator strength in the absence of doping $f_\mathrm{X}$~\cite{glazov2020optical}.
With the applied magnetic field the electron concentration difference in the two different valleys $\Delta n_e$ can be approximated as:
\begin{align}
    \Delta n_e \approx \frac{n_e\Delta_z}{2\epsilon_F}\frac{1}{1+e^{-\epsilon_F/(kT)}}~\label{eq:Zeeman}
\end{align}
where $\Delta_z=g_e\mu_B B$ is the electron Zeeman splitting, with the electron g-factor $g_e$ and Bohr magneton $\mu_B$, $\epsilon_F$ is Fermi energy, and $T$ is temperature.

Both observed effects can be modelled with a system of kinetic equations for the occupation numbers corresponding to A and B excitons $n_{XA}^{K(K')},n_{XB}^{K(K')}$ and trions $n_{T}^{K(K')}$ in the two valleys:
\begin{align}
    &\dot{n}_{XB}^{i}=P^i-\frac{n_{XB}^{i}}{\tau_{B,r}}-\frac{n_{XB}^{i}}{\tau_{intra}}-\frac{n_{XB}^{i}}{\tau_{inter}}+\frac{n_{XB}^{\bar{i}}-n_{XB}^{i}}{\tau_{dep,B}}\nonumber\\
    &\dot{n}_{XA}^{i}=-\frac{n_{XA}^{i}}{\tau_{A,r}}+\frac{n_{XA}^{\bar{i}}-n_{XA}^{i}}{\tau_{dep,A}}-\frac{n_{XA}^i}{\tau_{TF}^i}+\frac{n_{XB}^i}{\tau_{intra}}+\frac{n_{XB}^{\bar{i}}}{\tau_{inter}}\nonumber\\
    &\dot{n}_{T}^i=-\frac{n_{T}^{i}}{\tau_{T^i,r}}+\frac{n_{XA}^i}{\tau_{TF}^i}+\frac{n_{T}^{\bar{i}}-n_{T}^{i}}{\tau_{dep,T}} \label{eq:rateeq}
\end{align}
where $i,\bar{i}=K,K'$ is the valley index, $P$ is the pump rate into the specific valley,  $\tau_{[A,B,T^i],r}$ are the radiative lifetimes of A,B excitons and trions, $\tau_{dep,[A,B,T]}$ are the respective valley depolarization times, $\tau_{intra}$ and $\tau_{inter}$ are the characteristic times of the intra- and intervalley relaxation of B excitons to A excitons, and $\tau_{TF}$ is the trion formation time, which depends on the momentum distribution of the A excitons and thus on the temperature. 
Note that we account for the different trion radiative lifetimes due to the applied magnetic field. Since the trion formation rate depends on the residual electron concentration in the specific valley~\cite{glazov2020optical}, the applied magnetic field makes $\tau_{TF}$ valley dependent.

Using the proposed model, we calculate temperature dependencies of the trion DOCP and parameter $\rho$ by numerically finding the steady state solutions of Eqs.~\eqref{eq:rateeq} and evaluating the relative PL intensity as $I\propto n/\tau_{r}$.
The calculation results are shown in Fig.~\ref{fig:Temperature} with solid shaded curves and are in good agreement with the experimental data.
From the calculated temperature dependence of the $\rho$ parameter and Eq.~\eqref{eq:Zeeman} we estimate the value of the Fermi energy $\epsilon_F\approx 4$~meV and the Zeeman splitting $\Delta_z\approx 0.3$~meV.
This results in the estimated value of the effective magnetic field $B\approx 2.5$~T, which is much larger than the saturation magnetization field of BIG and thus may indicate the presence of the proximity effects in the heterostructure.


In summary, we have investigated circularly polarized photoluminescence of trions in monolayer MoSe$_2$ interfaced with ferrimagnetic BIG films.
The experimentally observed negative trion DOCP and contrast in co-polarized PL for the right and left circularly polarized excitation are explained with a model accounting for a an effective magnetic field due to the presence of BIG and valley-switching scattering from B to A excitons in MoSe$_2$.
Our results provide basic understanding of the mechanisms of trion polarization in such systems for the development of future devices based on the combination of electronic and valley degrees of freedom for information storage and processing.\\


The authors acknowledge funding from the Ministry of Education and Science of the Russian Federation through Megagrant No. 14.Y26.31.0015.
Optical measurements were funded by the Russian Science Foundation, project No. 19-72-00146.
V.K. acknowledges support by the Government of the Russian Federation through the ITMO Fellowship and Professorship Program.
The work of I.A.S. and I.V.I. on theoretical modelling of the described effects was funded by Russian Foundation for Basic Research (RFBR), according to the joint RFBR-DFG project No. 21-52-12038.
The work of P.O.K., V.I.B., and A.I.C. in terms of magnetic films and magnetic properties was supported by RSF project No. 21-12-00316.






\end{document}